\documentstyle[12pt,multicol,a4]{article}

\begin{document}
\input epsf

\begin{flushright}
DPNU-94-04\\
January 1994
\end{flushright}
\begin{center}
{\bf 
Critical Behavior of Black Hole Formation \\
in a Scalar Wave Collapse 
}
\end{center}
\vspace{0.5cm}
\begin{center}
  {\bf Y{\small oshimi} O{\small SHIRO},
    \footnote{e-mail: yoshiro@allegro.phys.nagoya-u.ac.jp }
       K{\small ouji} N{\small AKAMURA} 
    \footnote{e-mail: kouchan@allegro.phys.nagoya-u.ac.jp }
    }\\
  and {\bf A{\small kira} T{\small OMIMATSU} 
   \footnote{e-mail: c42615a@nucc.cc.nagoya-u.ac.jp}
    }
\end{center}
\vspace{0.3cm}

\begin{center}
  {\it Department of Physics, Nagoya University\\
    Chikusa-ku, Nagoya 464-01, Japan}
\end{center}
\vspace{0.5cm}

\baselineskip 2pc

We present a self-similar model of 
spherically symmetric collapse of a massless scalar field 
with a parameter $p$. The black hole formation is 
explicitly shown to occur only in the strong-field implosion of 
$p >1$. The field evolution in the critical limit $ p \rightarrow 1 $ 
is compared with numerical results found by Choptuik. 

\vspace{1cm}

Gravitational collapse is one of interesting 
phenomena in general relativity. 
In recent years, a large number of studies 
have been made on 
the dynamical behavior of a massless scalar 
field, in the context of 
spherically symmetric gravitational collapse. 
It has been shown that for a weak-field implosion
scalar waves bounce and disperse to infinity, 
while a strong-field implosion leads to black hole formation \cite{Goldwirth}. 
Furthermore, in numerical calculations, Choptuik \cite{Choptuik}
found an important behavior of a family of solutions which 
contain the parameter $p$ 
characterizing the strength of 
the scalar field: For black hole formation 
there exists the critical limit $p \rightarrow p^{*}$ 
to give the gravitational mass satisfying 
a power law $M_{BH} \propto  \mid p - p^{*} \mid^{\gamma}$ with 
a universal critical exponent $\gamma \simeq 0.37$.  
Such a critical behavior may be a general property of gravitational 
collapse, since imploding axisymmetric gravitational 
waves also show the similar behavior
\cite{Abrahams}.

In this letter our purpose is to present an analytical model 
of the spherically symmetric collapse of scalar waves. 
We solve the spherically symmetric Einstein equations 
coupled to a massless scalar field $\psi$  
to find the self-similar solution which has a critical parameter $p$.  
This model allows us to see explicitly 
the black hole formation and the scalar field dispersion. 
Our main concern is the field evolution in the critical limit 
$ p \rightarrow 1$, which should be compared with the numerical results.

Let us impose a self-similarity on the massless scalar field such that 
$\psi (u,v) = \psi (u/v) $, 
where $u$ and $v$ are retarded and advanced times respectively. 
The line element in the spherical symmetric double-null coordinates 
is given by 
\begin{equation}
ds^{2} = - h du dv + r^{2} d\Omega^{2}, \label{eqn:self-ds} 
\end{equation}
and we assume the metric to be $h(u,v) = h(u/v)$ and $r(u,v) = v f(u/v)$. 
For the energy-momentum tensor of the massless scalar field defined by 
\begin{equation}
T_{\mu\nu} = \frac{1}{4 \pi}
   ( \partial_{\mu} \psi \partial_{\nu} \psi - \frac{1}{2} g_{\mu\nu} 
\partial^{\alpha} \psi \partial_{\alpha} \psi   ), 
\end{equation}
the Einstein equations 
$G_{\mu\nu} = 8 \pi T_{\mu\nu}$ give the unique solution as follows, 
\begin{equation}
h = 1, 
\end{equation}
\begin{equation}
r^{2}  = \frac{1}{4} \{ (1-p^{2})v^{2} -2 vu +  u^{2} \}, \label{eqn:self-r}
\end{equation}
\begin{equation}
\psi = \pm \frac{1}{2} \log \frac{( 1 - p ) v - u }{ ( 1 + p ) v - u }, 
\end{equation}
where 
the constant $p$ can be chosen to be positive without loss of generality. 
In this letter we use units such that  $G=c=1$.  

The space-time is asymptotically flat at the past null infinity 
$ u \rightarrow -\infty$, where $r \simeq -u/2$. 
The scalar waves implode from the past null infinity with 
$\psi \simeq \pm p v /u \simeq \mp p v /r $. 
The parameter $p$ turns out to give the amplitude of the 
imploding scalar waves. 
This solution is useful to analyze the scalar wave collapse, 
if a pathological structure due to naked 
singularities is removed from the space-time.  
Because the Ricci scalar $R$ has the form 
\begin{equation}
R = \frac{ p^{2} uv }{2 r^{4} }, \label{eqn:singularity}
\end{equation}
the curvature singularity is located at the origin $r=0$. 
Note that the locus of $r=0$ is given by $u=(1-p)v$ and $u=(1+p)v$ in the 
regions $ v > 0 $ and $v<0$, respectively. 
The space-time structure is crucially dependent on the parameter $p$
(see Fig. 1). 
For $p>1$ the singularity becomes time-like in the region $v<0$ and 
spacelike in the region $v>0$, while for $0<p<1$ it always becomes timelike. 
(For $p=1$ exactly, the curvature singularity at $r=0$ in the 
region $v>0$ becomes null (i.e., u=0). We do not 
consider this case, and we will take the critical limit 
$p \rightarrow 1$ by keeping the condition $p>1$.)
The important point is that the apparent horizon exists only in the 
strong-field implosion of $p>1$. 
In fact we obtain 
the expansion $\theta$ for outgoing 
null rays given by 
\begin{equation}
 \theta = \frac{ \{ (1-p^{2}) v - u \} }{ 
                  \{(1-p) v  - u \} \{ (1+p)v  + u \} } 
\end{equation}
and the apparent horizon where $\theta=0$ 
is found to be $u= (1-p^{2}) v $. 
Thus the time-like singularity becomes naked, while the 
space-like one is hidden by the apparent horizon (see Fig. 1). 
This result will be due to the gravitational 
mass $M$ which can be locally defined in the spherically symmetric 
system as follows, 
\begin{equation}
M \equiv \frac{r}{2} ( 1 - g^{\mu\nu} \partial_{\mu} r \partial_{\nu}r ) 
        = - \frac{p^{2} uv }{8r}. 
   \label{eqn:def of mass}
\end{equation}
Near the singularity $u=(1-p)v$ in the region $v>0$, 
Eq. (\ref{eqn:def of mass}) is reduced to 
$M \simeq  p^{2} ( p - 1 ) v^{2} /r $. 
The mass turns out to be negative near the time-like singularity ($0<p<1$) 
and positive near the space-like one ($p>1$). 
Near another timelike singularity $u=(1+p)v$ in the region $v<0$, 
we obtain $M<0$ irrespective of $p$. 

In 2-dimensional gravity, 
the similar relation among the singularity, the local gravitational mass and 
the apparent horizon was pointed out by Hayward 
in the context of the cosmic censorship \cite{Hayward}. 
Note that the 2-dimensional metric can have the form $ds^2 = r^{-2} du dv $ 
with $r$ given by Eq. (\ref{eqn:self-r}) as a special case. 
This may suggest that the 2-dimensional dilation gravity works as a good 
model of the spherically symmetric 4-dimensional gravity. 
However, in the 4-dimensional space-time, 
the dynamical scattering of imploding scalar waves plays an important 
role for the evolution of the scalar field. The critical behaviors 
which will be discussed  in the following are missed in 
the 2-dimensional model.

It is clear that the self-similar solution is not applicable to the 
region $v<0$, where the naked timelike singularity $u=(1+p)v$ appears 
with $M<0$. For a realistic evolution we must avoid such a singularity 
to set some regular initial data. 
Fortunately, this is possible, since the solution requires that 
there is no energy flux across the $v=0$ line 
(i.e. outgoing wave $\psi_{u}$ equal to zero.) 
and the mass is equal to zero on it. 
We can replace the region $v<0$ by a flat space-time, which is 
smoothly matched to the region $v>0$ (see. Fig 2). 
For the weak-field implosion ($0<p<1$) shown in Fig. 2a, 
the region corresponding to $u>0$ and $v>0$ is also replaced by a flat 
metric, and the time-like singularity $u=(1-p)v$ is eliminated from the 
space-time. (Recall that $M=0$ on the $u=0$ line.) This is the case 
that the dynamical scattering of the imploding waves can induce 
a complete bounce to the future null infinity $v \rightarrow \infty$. 
For the strong-field 
implosion ($p>1$) shown in Fig. 2b, the space-like singularity 
$u=(1-p)v$ is hidden by apparent horizon. This means a black hole formation 
by imploding scalar waves. However, our self-similar model requires the 
imploding amplitude increasing like $\mid \psi \mid \sim v$ and 
fails to give an asymptotically flat region at the future null infinity 
$v \rightarrow \infty$. 
(The radius $r = p(p^{2} - 1 )^{1/2} v/2$ of the apparent 
horizon infinitely increases as $v \rightarrow \infty$.)
Therefore, for gravitational collapse of wave packets shown in Fig. 3, the 
self-similarity must break down in some region denoted by $v>v_{0}$, where 
outgoing waves dominate incoming ones. Nevertheless the discussions 
in the following suggest that the self-similar solution is approximately 
valid in a strong-field region of $0 \leq v < v_{0}$.

We have found the crucial difference of the space-time structure between 
the strong-field implosion ($p>1$) and the weak one ($0 <p<1$). 
Now let us consider the critical limit $ p \rightarrow 1$ 
for black hole formation. 
The numerical calculations of wave-packet collapse done by Choptuik claim 
that the black hole mass approaches zero in some critical limit. 
To discuss such a critical behavior for our self-similar model, 
we must estimate the gravitational mass $M_{h}$ on the apparent horizon 
instead of the Bondi mass which can be defined only at the future null 
infinity $v \rightarrow \infty$. 
>From Eq. (\ref{eqn:def of mass}) we obtain 
$M_{h}=p \sqrt{p^2 - 1 } v /4 $. It is interesting to compare 
$M_{h}$ with the initial mass 
$ M_{i} = p^{2} v /4 $ at the past null infinity $u \rightarrow -\infty$. 
The ratio 
\begin{equation}
\frac{M_{h}}{M_{i}} = \frac{\sqrt{p^{2} - 1}}{p}. 
                  \label{eqn:mass ratio}
\end{equation}
may be interpreted as the transmission coefficient of 
incoming waves propagating along a fixed $v$. For the strong-field limit 
$p >>1$ the amount of dynamical scattering which generates outgoing 
waves remains negligibly small before incoming waves reach the apparent 
horizon. 
For the critical limit $p \rightarrow 1$, however, almost all the 
incoming waves are scattered away, and a small portion of the generated 
outgoing waves can fall into the apparent horizon. 
(Recall that the surface $u=(1-p^{2})v$ becomes nearly null.) 
Thus the black hole mass $M_{h}$ estimated on the apparent horizon 
vanishes in the limit $p \rightarrow 1$, which gives the power law 
\begin{equation}
M_{h} \simeq (\frac{p-1}{8})^{\frac{1}{2}} v. \label{eqn:critical}
\end{equation}
This is slightly different from the power law with the 
critical exponent $\gamma^{*} \simeq 0.37$, which was 
numerically evaluated by Choptuik at the future null infinity 
$ v \rightarrow \infty$. We wish to emphasize that Eq. (\ref{eqn:critical}) 
exhibits a mass evolution in some dynamical stage of gravitational collapse 
instead of the asymptotic stage $v \rightarrow \infty $. 
If the self-similarity approximately holds in the stage 
$0<v<v_{0}$ even for the wave-packet collapse shown in Fig. 3, the 
final mass will be given by Eq. (\ref{eqn:critical}) with 
$v_{0}$ in place of $v$. Then, to obtain the critical exponent 
$\gamma^{*} \simeq 0.37$, the advanced time $v_{0}$ should increase like 
$v_{0} \propto \mid p -1 \mid^{-0.13}$ in the limit $p \rightarrow 1$. 
For the near-critical evolution the self-similar stage may continue for 
a long time. Though this problem of the critical exponent remains 
unclear within the framework of the present analysis, we can point out 
another critical behavior in relation to the numerical result. 

Now let us introduce the time coordinate $t$ defined by 
\begin{equation}
\log(t/r) = -  \int [
     \frac{ \sqrt{p^{2}X^{2} +4}- p^{2}X }{
    2 - X \sqrt{p^{2}X^{2} +4} +  p^{2} X^{2} } 
                  ] dX , 
\label{eqn:def t} 
\end{equation}
where $ X \equiv  v / r $.  The line element can be rewritten 
into the diagonal form 
\begin{equation}
ds^{2} = - \alpha (X) dt^{2} + \beta (X) dr^{2} 
     +r^{2}d\Omega^{2},
\end{equation}
where the metric coefficients $\alpha(X)$ and $\beta(X)$ are 
\begin{eqnarray}
 \alpha(X) &=& (1-\frac{ p^{2} X }{ \sqrt{ p^{2}X^{2}+4 } } ) 
     [
     \frac{ 2 - X \sqrt{p^{2}X^{2} +4} +  p^{2} X^{2} }{ 
             \sqrt{p^{2}X^{2} +4}- p^{2}X } 
                  ]^{2} \times
      \nonumber\\
  & & 
 \exp ( 2 \int [
     \frac{ \sqrt{p^{2}X^{2} +4}- p^{2}X }{
    2 - X \sqrt{p^{2}X^{2} +4} +  p^{2} X^{2} } 
                  ] dX ),
      \label{eqn:alpha}  \\
 \beta(X) &=&  
        \frac{4}{\sqrt{p^{2}X^{2} + 4} (\sqrt{p^{2} X^{2} +4} -p^{2}X )}, 
       \label{eqn:beta}
\end{eqnarray}
The scalar field $\psi$ also depends on $X$ only, 
\begin{equation}
\psi = \pm \log \frac{1}{4} ( \sqrt{ p^{2} X^{2} + 4 } + p X ). 
            \label{eqn:psi-x}
\end{equation}
We consider the field evolution measured by the coordinates $t$ and $r$, 
which are related to the variable $X$ through Eq. (\ref{eqn:def t}). 
The variable $X$ satisfies $X^{2} < X^{2}_{h} \equiv 4/ p^{2} ( p^{2} -1)$ 
outside of the apparent horizon. 
Then, in the critical limit $p \rightarrow 1$, 
there exists the strong-field region $1 << X^{2} << X^{2}_{h}$ in a 
neighborhood of $r=0$, where Eq. (\ref{eqn:def t}) gives the logarithmic 
relation 
\begin{equation}
\frac{1}{(p-1)X^{2}} = \log t - \log r.  \label{eqn:log dep}
\end{equation}
Such a logarithmic evolution was also observed in the numerical 
calculation of wave-packet collapse with a scaling relation. 
Note that the time ``t" proportional to the 
centoral proper time which defined by 
$ T \equiv \int \sqrt{ \mid g_{tt}(t,0) \mid } \, dt \simeq 
\sqrt{ \mid p-1 \mid } \, t $.  
For the self-similar solution the scaling relation 
$\psi(\log r -\triangle , \log t - \triangle ) = 
 \psi( \log r, \log t)$ 
clearly hold with a continuous parameter $\triangle$. 
For the wave-packet collapse, however, the scaling parameter $\triangle$ 
was found to be restricted to some discrete values $n \triangle^{*}$ 
($\triangle^{*} \simeq 3.4$ and $n = 1,2,3, \dots $). 
This difference of the scaling will not be crucial, if we consider 
the region where both $\log r$ and $\log t$ are much larger than 
$\triangle^{*}$. 
The apparent horizon can be arbitrarily close to $r=0$ in the 
limit $p \rightarrow 1$. Then the relation 
$\log r >> \triangle^{*}$ is satisfied in the strong-field region 
outside the apparent horizon near $r=0$, where the spacing of the 
discrete values $n\triangle^{*}$ becomes unimportant to recover 
approximately the self-similarity.

In summary, our self-similar model can give a simple picture of 
black hole formation due to imploding scalar waves with the transmission 
coefficient (\ref{eqn:mass ratio}). 
We obtain the critical behaviors which exhibit the mass evolution 
(\ref{eqn:critical}) on the apparent horizon and the logarithmic 
evolution (\ref{eqn:log dep}) in a neighborhood of $r=0$. 
The scaling  relation means that the self-similarity may appear even 
in wave-packet collapse with the parameter $p$ sufficiently close 
to the critical value and can represent a dynamical growth of the 
apparent horizon, though the critical exponent $\gamma^{*}$ and 
the scaling constant $\triangle^{*}$ are missed. 
It is necessary to check the self-similar critical behaviors in 
numerical calculations of wave-packet collapse.

\begin{center}

Acknowledgments

\end{center}
This work is partially supported by the Grant-in-Aid for 
Scientific Research from the Ministry of Education, 
Science and Culture (04640268).

\newpage

\begin{center}
Figures
\end{center}

\begin{multicols}{2}
    \epsfxsize=6cm \epsfbox[45 0 503 748]{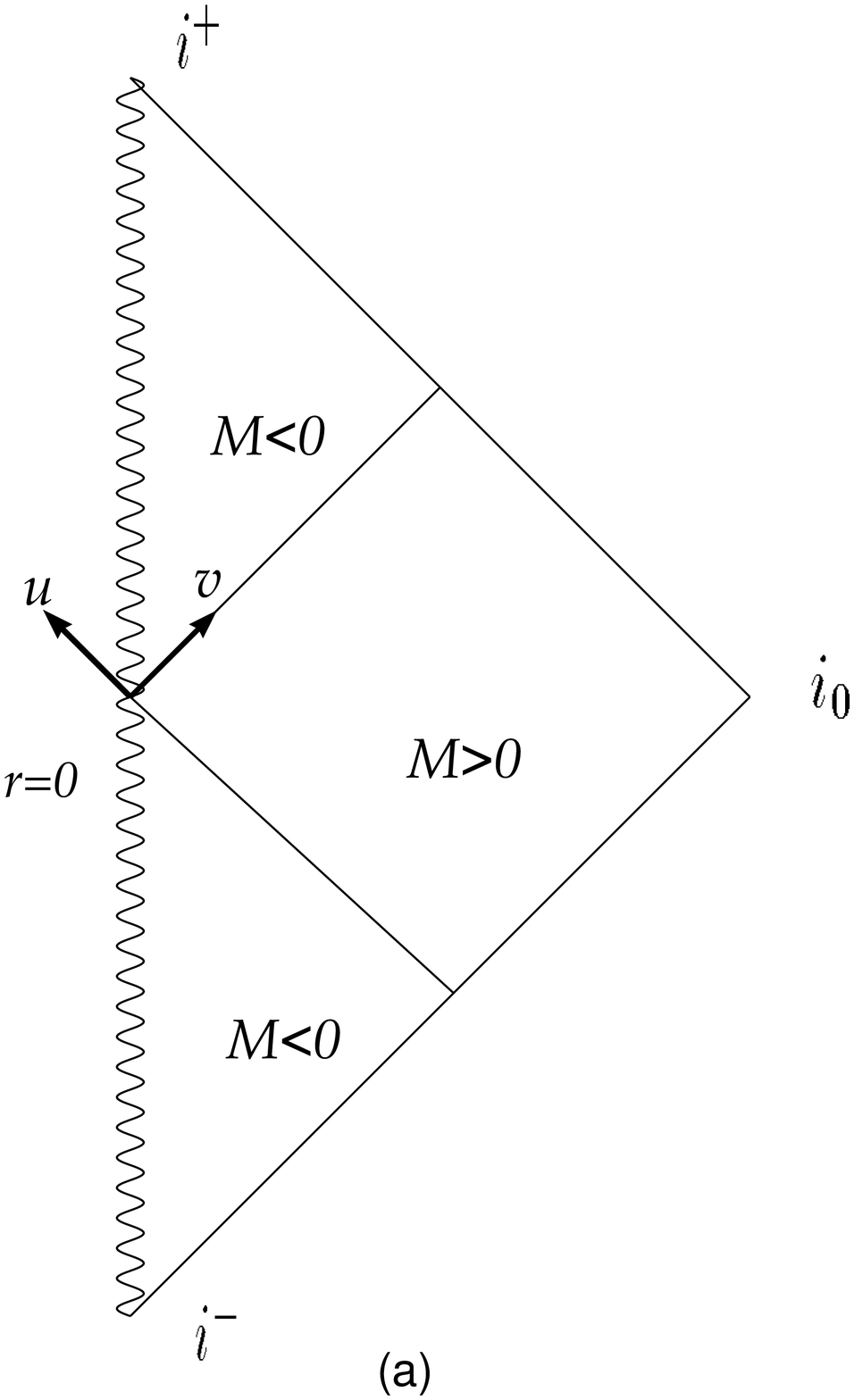}
    \epsfxsize=6cm \epsfbox[45 0 503 748]{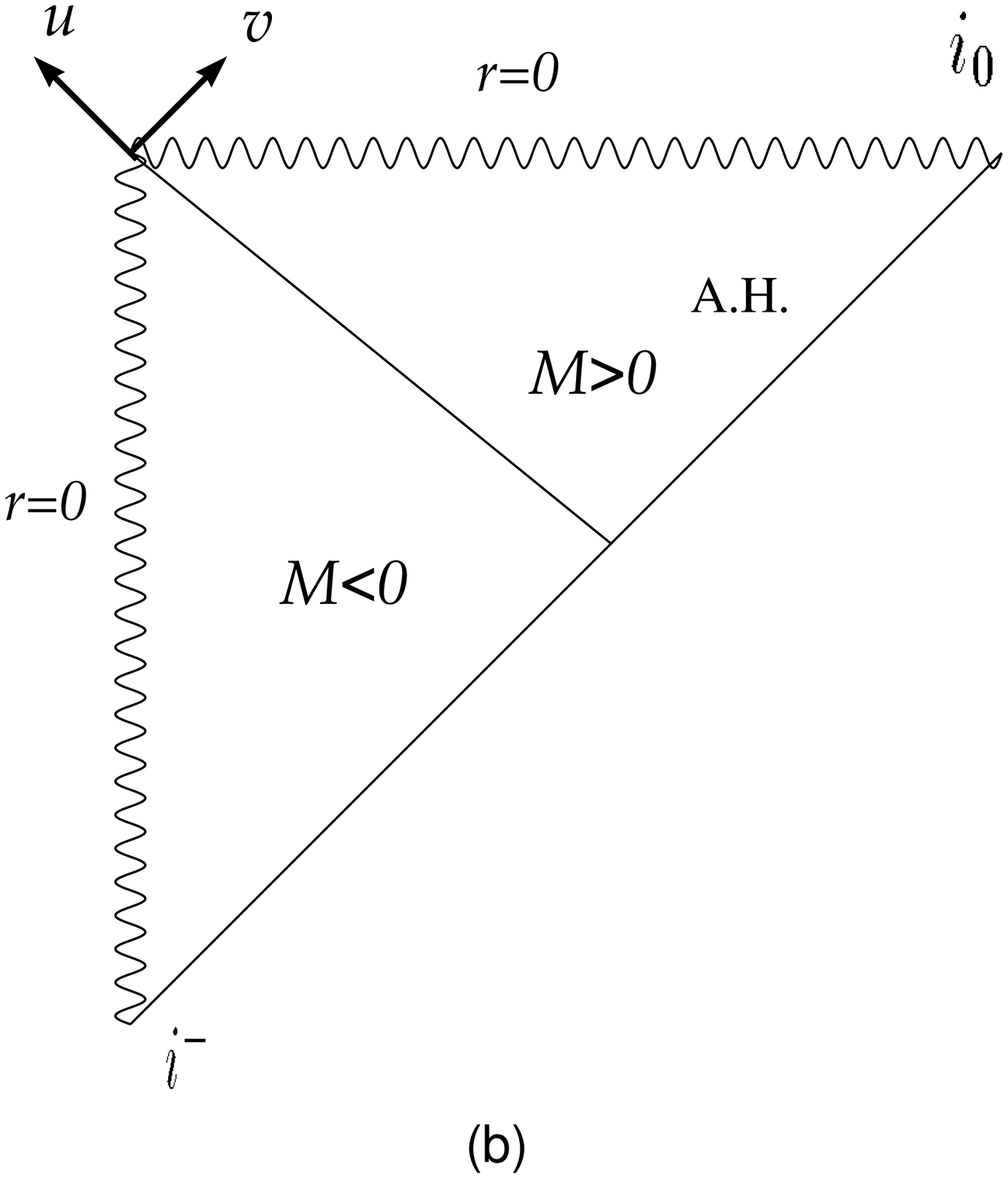}
\end{multicols}

\begin{center}
  Fig. 1. 
\end{center}

Penrose diagram for the self-similar solutions
covering the whole space-time. The scalar waves implode 
from the past null infinity $\cal J^{-}$ with the 
amplitude denoted by the parameter $p$: 
(a) For the weak-field implosion $0<p<1$  the 
time-like singularity appears in the negative mass regions of 
$uv >0$. 
(b) For the strong-field implosion $p>1$ the
apparent horizon (A.H.) is formed to surround the space-like singularity 
in the positive mass region. 
This space-time has no future null infinity $\cal J^{+}$
at $ v \rightarrow \infty $. 

\newpage

\vspace{0.5cm}

\begin{multicols}{2}
    \epsfxsize=6cm \epsfbox[32 0 475 748]{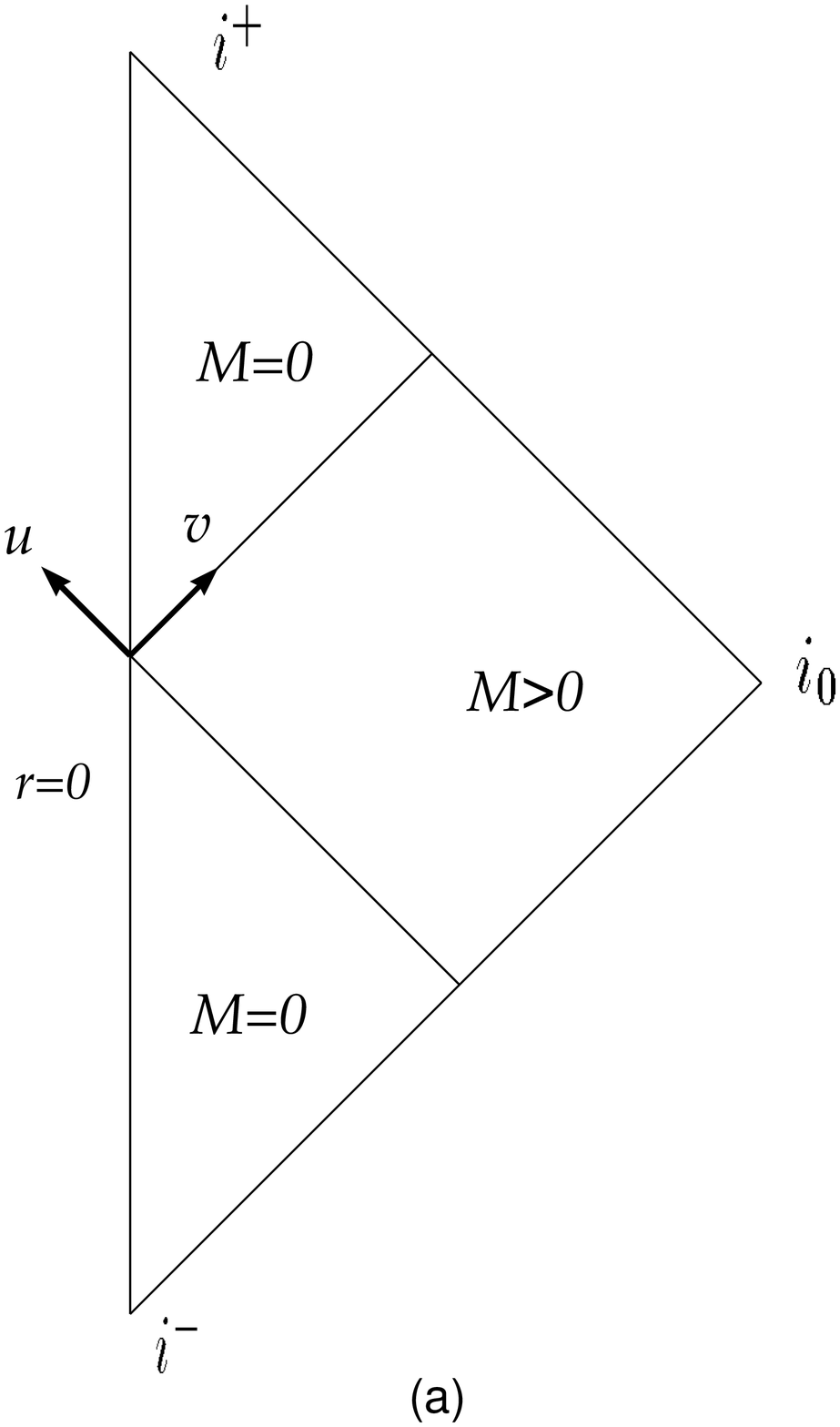}
    \epsfxsize=6cm \epsfbox[32 0 475 748]{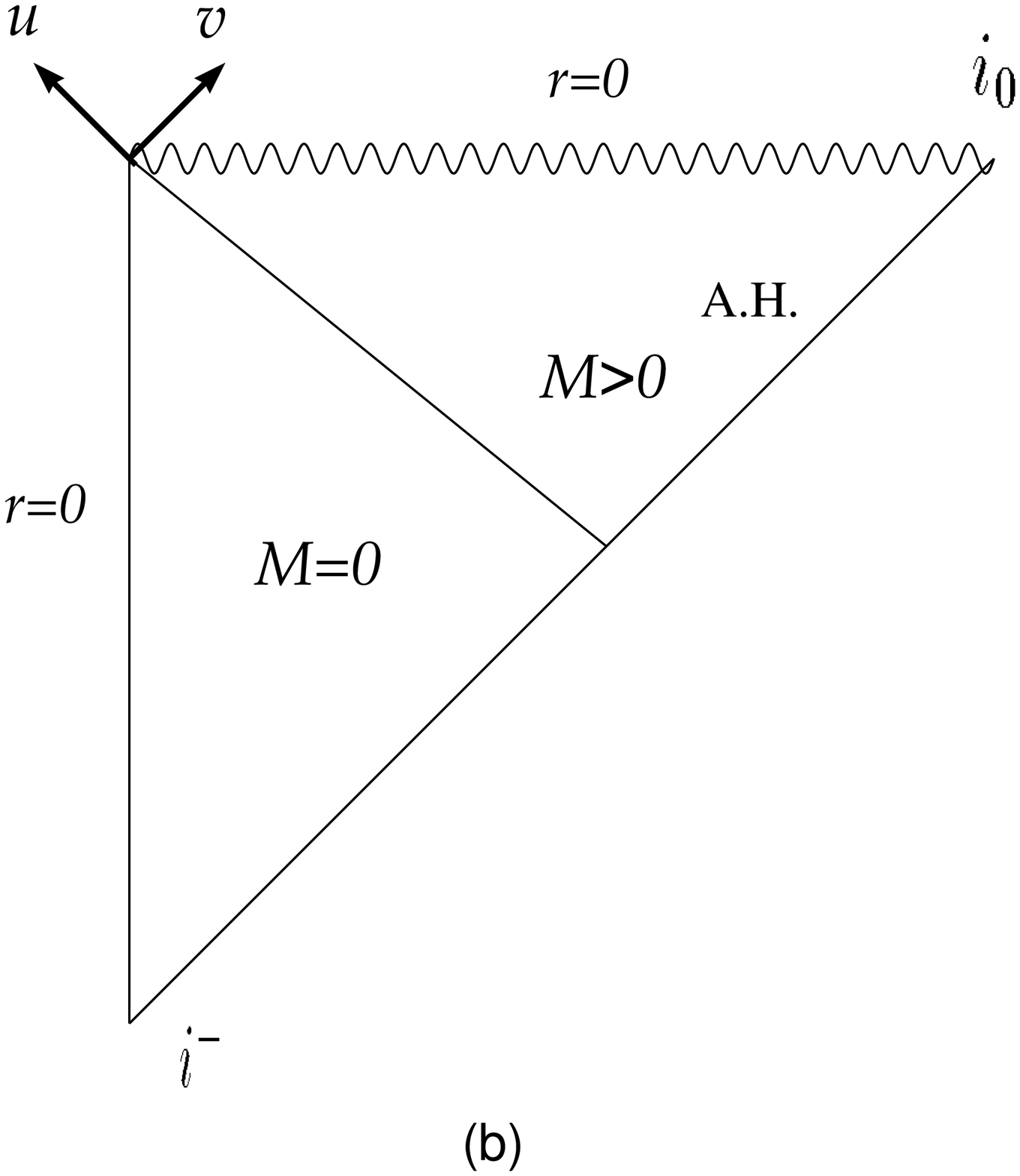}
\end{multicols}

\begin{center}
  Fig. 2. 
\end{center}

Penrose diagram for the self-similar solution, in which the 
negative mass region are replaced by the flat space-time with 
$M=0$. 

(a) For $0<p<1$ all the imploding waves bounce and disperse to 
the future null infinity $\cal J^{+}$. 
(b) For $p>1$ we obtain a model of black hole formation due to the 
scalar wave collapse, though the mass infinitely increases as 
$v \rightarrow \infty$. 

\vspace{0.5cm}

\epsfxsize=9cm \epsfbox[-100 36 324 548]{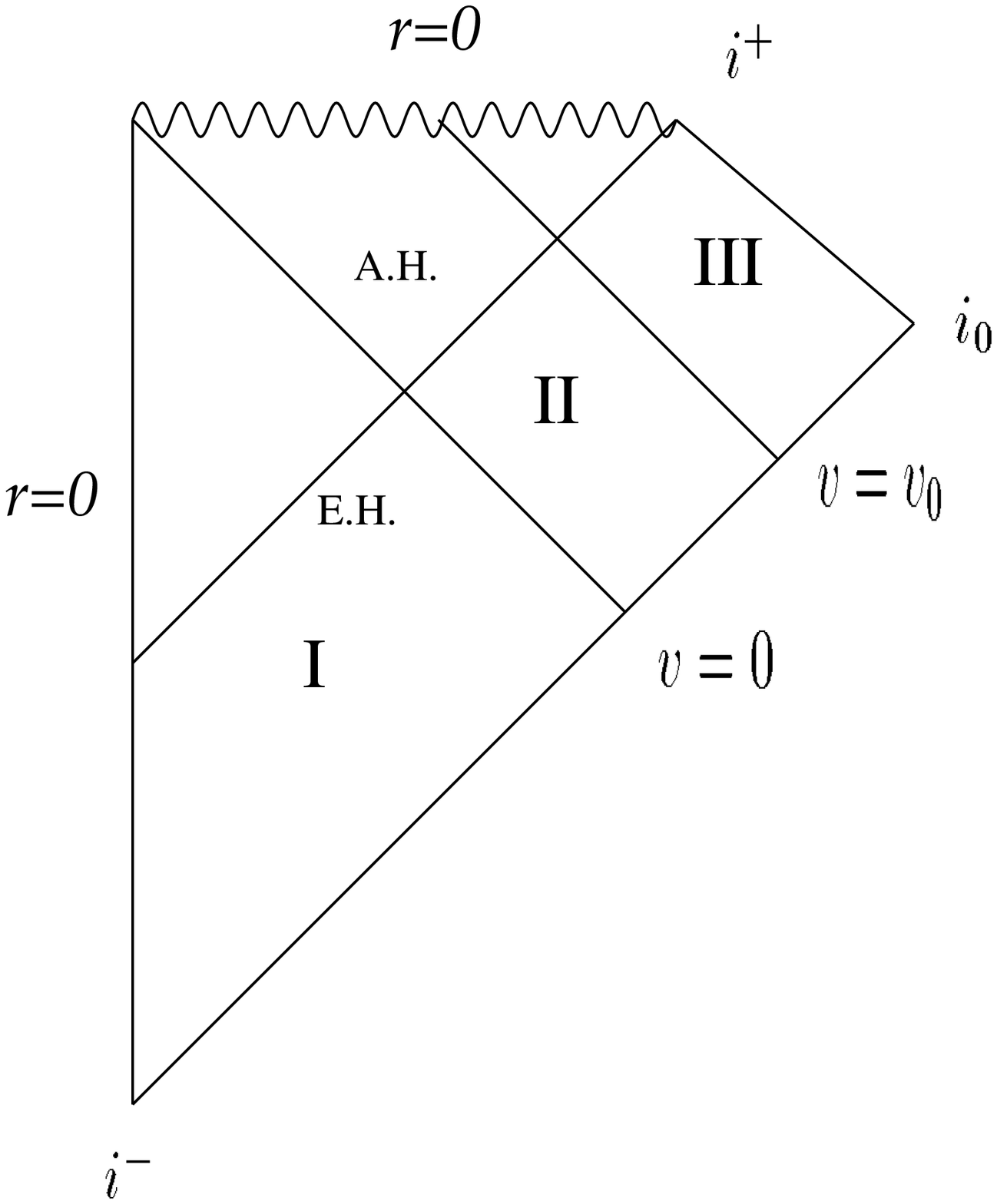}

\begin{center}
  Fig. 3
\end{center}

Penrose diagram for black hole formation due to wave-packet collapse. 
The region I ($v <0$) is the flat space-time, and in the region III 
($v > v_{0}$) outgoing waves dominate incoming ones. The self-similarity 
will remain valid in the region II ($ 0<v<v_{0}$).

\end{document}